\documentclass[epj]{webofc}
\usepackage[varg]{txfonts}

\newcommand{\mathd}{\mathrm{d}}
\newcommand{\mathi}{\mathrm{i}}
\newcommand{\mathe}{\mathrm{e}}

\woctitle{QCD@Work 2014}
\begin{document}
\title{Fuzzy bags, Polyakov loop and gauge/string duality}
%
%
\author{Fen Zuo\inst{1}\fnsep\thanks{\email{zuofen@hust.edu.cn}}
}
\institute{School of Physics, Huazhong University of Science and Technology, Wuhan 430074, China
          }

\abstract{%
  Confinement in SU($N$) gauge theory is due to the linear potential between colored objects. At short distances, the linear contribution could be considered as the quadratic correction to the leading Coulomb term. Recent lattice data show that such quadratic corrections also appear in the deconfined phase, in both the thermal quantities and the Polyakov loop. These contributions are studied systematically employing the gauge/string duality. "Confinement" in ${\mathcal N}=4$ SU($N$) Super Yang-Mills~(SYM) theory could be achieved kinematically  when the theory is defined on a compact space manifold. In the large-$N$ limit, deconfinement of ${\mathcal N}=4$ SYM on $\mathbb{S}^3$ at strong coupling is dual to the Hawking-Page phase transition in the global Anti-de Sitter spacetime. Meantime, all the thermal quantities and the Polyakov loop achieve significant quadratic contributions. Similar results can also be obtained at weak coupling. However, when confinement is induced dynamically through the local dilaton field in the gravity-dilaton system, these contributions can not be generated consistently. This is in accordance with the fact that there is no dimension-2 gauge-invariant operator in the boundary gauge theory. Based on these results, we suspect that quadratic corrections, and also confinement, should be due to global or non-local effects in the bulk spacetime.}
\maketitle
\section{Fuzzy bags and Polyakov loop}
Lattice data for the heavy quark potential can be well described by the Cornell form, which is the sum of the linear term and the Coulomb one. The linear term is responsible for confinement at large distance. At short distance $r\to 0$, the Coulomb term makes the dominant contribution and the linear one gives a small $r^2$ correction~\cite{Akhoury:1997by}. In the momentum picture, this means a correction with the behavior $1/Q^2$ at large $Q$. However, since no dimension-2 gauge-invariant operator exists, it is difficult to generate such contributions through the operator product expansion.
Interestingly, recent lattice data shows that such quadratic corrections also exist above the deconfinement phase transition.
It is found that the lattice result for the trace anomaly $\Delta$ in SU($3$) gauge theory, normalized in unit of $T^4$, is well described by a $1/T^2$ term. Such an observation is made manifest by plotting the ratio $\Delta/T^2$ versus $T$ in the deconfined phase~\cite{Pisarski:2006yk}. In a large temperature region $T_c\lesssim T\lesssim 4T_c$, the ratio takes nearly a constant value. Such behavior is further confirmed in SU($N$) gauge theory for various $N$~\cite{Panero:2009tv}.
Based on such an observation, a fuzzy bag model is proposed, in which the pressure takes the form
\begin{equation}
p_{\rm{QCD}}\approx f_{\rm{pert}} T^4-B_{\rm{fuzzy}}T^2-B_{\rm{MIT}}+...\label{eq.FB}
\end{equation}
In addition to the ideal term and the bag constant contribution, a $T^2$-term is introduced to describe the novel behavior of the trace anomaly. Such a term indicates that confinement should not be realized by a rigid MIT bag. A realistic bag must have fluctuant thickness which induces the temperature dependent contribution. Again we see that the quadractic term is intimated related to the confining mechanism.

At finite temperature, one defines the Polyakov loop as the Wilson loop along the Euclidean time circle. A nonzero value of the Polyakov loop signals the breaking of the center of the gauge group. It is thus taken as the order parameter of the deconfinement phase transition. One would naturally expect that it also achieves significant quadractic corrections in the deconfined phase. It is first pointed out in \cite{Megias:2005ve} that the logarithm of the Polyakov loop in SU(3) gauge theory is dominated by a $1/T^2$ term. Recently, similar pattern has also been observed in SU($4$) and SU($5$) gauge theory~\cite{Mykkanen:2012ri}.

\section{${\mathcal N}=4$ Super Yang-Mills on $\mathbb{R}^3$ and $\mathbb{S}^3$}
In the large-$N$ limit, ${\mathcal N}=4$ SU($N$) Super Yang-Mills~(SYM) gauge theory is conjectured to be dual to Type IIB string theory~\cite{Maldacena:1997re}, living on 10 dimensional spacetime given by the product of 5D  Anti-de Sitter~(AdS) and a 5 sphere. The duality involves a part of the total AdS$_5$ described by~\cite{Maldacena:1997re}
\begin{equation}
 \mathd s^2 =\frac{r^2}{L^2}-\text{d} t^2 +\left\{\frac{r^2}{L^2}\right\}^{-1}\mathd r^2+\frac{r^2}{L^2}\mathd \mathbf{x}^2, r\ge0.\label{eq.AdS1}
\end{equation}
The gauge theory lives on the boundary $r=\infty$, which could be thought of as 4 dimensional Minkowski spacetime~\cite{Witten:1998qj}.
Due to conformal invariance, no bound states exist and the theory is unconfined.

How about the other regions of the AdS spacetime? The whole AdS space is described by~\cite{Hawking:1982dh,Witten:1998zw}
\begin{equation}
\mathd s^2=-\left(1+\frac{r^2}{L^2}\right)\mathd t^2+\left(1+\frac{r^2}{L^2}\right)^{-1}\text{d} r^2+r^2\mathd \Omega_3^2, \label{eq.AdS2}
\end{equation}
where $t$ is periodically identified with period $2\pi L$, and $\mathd \Omega_3^2$ the metric on a 3 sphere. Now the boundary of the space, also given by $r=\infty$, has the topology $\mathbb{S}^1\times \mathbb{S}^3$. The Gauss constraint on the compact manifold forced the colored states to disappear. One obtains kinematically "confinement". Comparing with eq.~(\ref{eq.AdS1}), one finds that in eq.~(\ref{eq.AdS2}) there is an additional factor $1$ in the warp factors. According to the dictionary of the duality~\cite{Witten:1998qj}, it induces quadratic correction in the boundary theory. It is interesting that "confinement" comes together with quadractic corrections, though it is not dynamical.

One may expect that these quadratic corrections also appear at finite temperature. Thermal states on AdS space can be constructed by substituting $\tau=\mathi t$, and periodically identified $\tau$ with period $\beta=T^{-1}$. Except the above static AdS solution~(\ref{eq.AdS2}), the Einstein equations also admit black hole solutions. It can be obtained from the usual Schwarzschild black hole by adding the contribution from the cosmological constant $\Lambda=-3/(8L^2)$:
\begin{equation}
\mathd s^2=-\left(1-\frac{\omega_{4}M}{r^2}+\frac{r^2}{L^2}\right)\mathd t^2+\left(1-\frac{\omega_{4}M}{r^{2}}+\frac{r^2}{L^2}\right)^{-1}\mathd r^2+r^2\mathd \Omega_3^2,\label{eq.AdS-BH1}
\end{equation}
where $M$ is the mass of the black hole, and $\omega_{4}=\frac{16\pi G_{5}}{{\rm{Vol}}(S^3)3}$. When $M\to 0$ the black hole disappears and one recovers (\ref{eq.AdS2}). The black hole horizon and the Hawking temperature are given by
\begin{equation}
M=\frac{1}{\omega_{4}}\left[r_+^{2}+\frac{r_+^{4}}{L^2}\right],~~~T_H=\frac{2r_+^2+L^2}{2\pi L^2r_+}.
\end{equation}
So for $T_H> T_{\rm{min}}=\frac{\sqrt{2}}{\pi L}$ there will be two solutions with different $r_+$. At $T_{\rm{min}}$ the two solutions coincide. The dominant phase is give by the solution with small free energy. Explicit calculation shows that the big black hole always dominates over the small one. Comparing to the static solution (\ref{eq.AdS2}), the situation changes with temperature. It turns out that the thermal solution (\ref{eq.AdS2}) is preferred at low temperature, while the big black hole makes the dominant contribution at high temperature. The transition, first pointed out by Hawking and Page, occurs at the temperature $T_c=\frac{3}{2\pi L}$. The free energy of the black hole solution is of order $N^2$, indicating that the high temperature phase is unconfined. Therefore the Hawking-Page transition corresponds to deconfinement in the boundary theory. Similar phase transition at weak coupling has been found in~\cite{Sundborg:1999ue}, known as the Hagedorn transition. One could further calculate the thermal quantities in the deconfined phase, and formally define the "trace anomaly" $\Delta=\epsilon-3p$. The result is~\cite{Zuo:2014vga}
\begin{equation}
\Delta/T_H^4=\frac{\pi^2N^2}{12}\frac{T_c^2}{T_H^2}\left[1+\sqrt{1-\frac{8}{9}\frac{T_c^2}{T_H^2}}\right]^2,\label{eq.D/p}
\end{equation}
which is indeed dominated by $1/T_H^2$ term. Similar results at weak coupling can be obtained following the formalism in~\cite{Aharony:2003sx}.

It would be interesting to compare the above results with those when the boundary space is $\mathbb{R}^3$. The corresponding black hole solution can be obtained by taking the large-$M$ limit of eq.~(\ref{eq.AdS-BH1}). Absorbing the divergent quantities by redefining the coordinates, one arrives at the metric
\begin{equation}
\mathd s^2 =-\frac{r^2}{L^2}\left[1-\left(\frac{r_0}{r}\right)^{4}\right]\mathd t^2 +\left\{\frac{r^2}{L^2}\left[1-\left(\frac{r_0}{r}\right)^{4}\right]\right\}^{-1}\mathd r^2+\frac{r^2}{L^2}\mathd \mathbf{x}^2.\label{eq.AdS-BH2}
\end{equation}
 Here the horizon is $r_0$ and the temperature is given by $T_H=\frac{1}{\pi L^2} r_0$. Due to conformal invariance, making the radius of $\mathbb{S}^3$ infinite is equivalent to taking the large temperature limit. There the theory on $\mathbb{R}^3$ has a unique nonzero temperature phase, the deconfined phase, described by the above black hole solution. Conformal invariance also forces the trace anomaly to vanish.

The deconfinement transition could be further investigated through the behavior of the Polyakov loop.
 According to the duality, it could be derived from the area of the string worldsheet ending on the time circle $\mathbb{S}^1$. No such worldsheet exists for the thermal solution (\ref{eq.AdS2}) due to the topology $\mathbb{S}^1\times\mathbb{S}^4$~\cite{Witten:1998zw}. The Polyakov loop takes zero expectation value, corresponding to the confining feature. The topology of the black hole solution (\ref{eq.AdS-BH1}) is $\mathbb{R}^2\times\mathbb{S}^3$. The string worldsheet ending on $\mathbb{S}^1$ simply expands the whole $\mathbb{R}^2$ part. Calculating the area of such a worldsheet then gives the quark free energy, and further the Polyakov loop. However, the area of such a worldsheet is divergent. The divergence persists in the large-$T$ limit, which can be seen from the metric (\ref{eq.AdS-BH2}). 
 To achieve a finite result for the Polyakov loop, we choose to subtract the area of the string worldsheet in the background (\ref{eq.AdS-BH2}) from that in (\ref{eq.AdS-BH1}). In this way, the required information of the Polyakov loop related to confinement is extracted. With this substraction, the Polyakov loop value can be obtained~\cite{Zuo:2014vga}
 \begin{equation}
 -2\log {\mathcal L}^R=\frac{\lambda}{2}\left[1-\sqrt{1-\frac{8}{9}\frac{T_c^2}{T_H^2}}\right],
 \end{equation}
where $\lambda\equiv g_{\rm{YM}}^2 N=4\pi g_sN$. The finite value of the Polyakov loop again shows that the high temperature phase is deconfined. The above formula is dominated by a $1/T_H^2$ term, which is just the observation in lattice simulation. It will be interesting to calculate the Polyakov loop at weak coupling using the technique in \cite{Aharony:2003sx}.

\section{The gravity-dilaton system}
It would be interesting to generalize the above discussion to realistic theories, where confinement is generated dynamically. A concrete criterion for confinement is the area law of large Wilson loop, which is a direct consequence of the linear potential. From gauge/string duality, the expectation value of the Wilson loop is calculated from the area of the string worldsheet ending on the loop. When will the area of the string worldsheet be proportional to the loop area on the boundary? The answer is when the infrared part of the bulk spacetime is cut off~\cite{Witten:1998zw}. In this case the string worldsheet is forced to dwell at the cutoff. This part makes the dominant contribution and is proportional to the loop area at the boundary. The resulting model exhibits linear potential and confinement. An explicit model is given by the D4 brane construction~\cite{Witten:1998zw}.

\subsection{Hard wall and soft wall}

Phenomenologically, one can introduce a cutoff by hand into the AdS background (\ref{eq.AdS1}), which is called a hard wall. 
The cutoff breaks conformal invariance and induces bound states in the model.
However, the mass pattern of the excited states, $m_n\propto n$, is in contradiction to the observed Regge trajectory $m_n\propto\sqrt{n}$. When the cutoff $z_0$ is hidden inside the black hole horizon, the model jumps to the deconfined phase~\cite{Herzog:2006ra}. A simple calculation leads to the pressure in the black hole phase
\begin{equation}
p\propto T^4-\frac{2}{\pi^4 z_0^4}+{\cal O}(T^{-2}).
\end{equation}
In addition to the leading ideal term, a constant term appears similar to the bag model. In this sense, the hard wall could be thought of as a holographic bag model.

How to reproduce the correct mass spectra and the fuzzy term? A nontrivial dilaton field, $\Phi(z)\sim \Lambda^2 z^2$, is introduced to the AdS background. 
Alternatively, one could introduce an exponential factor $\exp\{-2\Lambda^2z^2\}$ into the metric to suppress the contribution from the infrared region ($z\to \infty$). 
Since the infrared region is only suppressed rather that cut off, the model is called soft wall. The resulting spectra turns out to be Regge like, with the Regge slope $4\Lambda^2$. However, a direct calculation of the heavy quark potential shows that the model is not really confined~\cite{Zuo:2009dz}. 
To achieve confinement, the dilaton must behave as $\Phi(z)\sim \Lambda^2 z^2$. Accordingly, the string frame metric needs to be enhanced by a factor $\exp\{2\Lambda^2z^2\}$ in the infrared. With such a factor, the string frame metric exhibits a minimum at $z_m=1/(\sqrt{2}\Lambda)$, which plays the same role as the cut off in the hard wall model. The string worldsheet prefers to dwell at the minimum, resulting in area law of the Wilson loop. The deconfinement transition takes place when the minimum coincides with the black hole horizon, at the temperature $T_c=\sqrt{2}\Lambda/\pi$~\cite{Andreev:2006eh}.
As the temperature increases above $T_c$, the minimum is hidden inside the horizon and the model unconfines. The pressure in the deconfined phase can be estimated to be
\begin{equation}
p\propto T^4(1-\frac{4\Lambda^2}{\pi^2 T^2}+{\cal O}(T^{-4})).
\end{equation}
One explicitly realizes the fuzzy bag model (\ref{eq.FB}) in the holographic approach. However, the model is not dynamically generated from a concrete theory, and thus contains some inconsistence.

\subsection{The gravity-dilaton construction}
To dynamically generate the soft wall, the coupling between the gravity and the dilaton should be considered. The general action takes the form
\begin{equation}\label{eq:S5}
S_5 = \frac{1}{2 \kappa^2} \int_M \!\! d^5x \; \sqrt{-g}\, \left (R - V(\phi) - \frac{1}{2} (\partial \phi)^2 \right),
\end{equation}
 plus possible boundary terms. Due to different form of the dilaton potential, the solutions exhibit different infrared behavior. At zero temperature, these solutions can be classified with the different infrared behavior~\cite{Gursoy:2007er}:
\begin{equation}
ds^2_0 \to \mathe^{- C z^{\alpha}}\!\!\left(dz^2+dx_4^2\right).
\end{equation}
\begin{equation}
\lambda_0 \equiv\mathe^{\sqrt{3/8}\phi}\to \mathe^{(3C/2)~ z^{\alpha}}\!\!z^{\frac{3}{4}(\alpha-1)}.
\end{equation}
The theory is confining if and only if $\alpha\ge 1$. In this case the theory exhibits discrete spectrum, and the spectrum pattern is determined by $\alpha$: $m_n^2\sim n^{2(\alpha-1)/\alpha}$. Choosing $\alpha=2$, one realizes explicitly the soft wall model. In the limit $\alpha\to \infty$, the spectrum turns to be $m_n\propto n$, as in the hard wall.

At finite temperature, there exists two kinds of solutions~\cite{Gursoy:2008bu}, similar as the thermal solutions in global AdS discussed in the previous section. The first kind is the direct thermalization of the zero-temperature background with $\tau=\mathi t$. Another kind of solutions  always contains the black hole, for any value of $\alpha$. If $\alpha<1$, the black hole solution dominates for the whole temperature region. The theory is always in the deconfined phase, similar as ${\mathcal N}=4$ SYM in flat spacetime. When $\alpha\ge1$, the black hole solutions exist only above a minimum temperature region $T_{\rm{min}}$. A phase transition occurs, from the thermal gas solution at low temperature to the black hole solution at high temperature. The phase transition is first order for $\alpha>1$, and second order for $\alpha =1$. It is very similar to the Hawking-Page transition discussed in the previous section, and inherits many properties there. The transition describes quite well the lattice data for the thermodynamic quantities~\cite{Gubser:2008yx,Gursoy:2008bu}. It is recently proposed that $\alpha$ could be directly related to the power of the thermal corrections~\cite{Caselle:2011mn}. However, we will show that this is not true.

\subsection{Power corrections in the gravity-dilaton system}
Since confinement and deconfinement phase transition can be implemented with proper dilaton potential, one would ask if those thermal power corrections are present together with confinement. 
To extract different power terms, we have to solve the theory asymptotically in the ultraviolet region. It is convenient to use the following metric form~\cite{Hohler:2009tv}
\begin{equation} \label{eq:ansatz1}
ds^2 = \frac{1}{z^2}\left(-f(z) dt^2 + d\vec{x}^2\right) + e^{2 g(z)} \frac{dz^2}{z^2 f(z)}.
\end{equation}
Then the metric function $g(z)$ satisfies a very simple equation
\begin{equation}
\dot{g} = -\frac{1}{6}\, \dot{\phi}^2, ~~~~\dot{y}\equiv z\frac{\mathd}{\mathd z} y. \nonumber
\end{equation}
This equation shows clearly that the metric achieves corrections of $\phi$ only at even orders. At zero order of $\phi$, one gets simply the AdS black hole (\ref{eq.AdS-BH2}). At linear order, the dilaton in such a background can be solved~\cite{Cherman:2009tw}
\begin{equation}
\phi(z) = \phi_H \, {}_2F_1\left(1-\frac{\Delta_+}{4},\,\frac{\Delta_+}{4},\,1,\, 1- \frac{z^4}{z_H^4}\right),\nonumber
\end{equation}
where $\Delta_+$ is the dimension of the operator dual to $\phi$, $z_H$ is the black hole horizon and $\phi_H$ is the horizon value of $\phi$. Substituting this into the equations, one obtains the metric up to $\phi_H^2$. With these solutions, all the physical quantities can be expanded in powers of $\phi_H$. In particular, the leading power terms in the trace anomaly and the Polyakov loop ${\mathcal L}$ are~\cite{Zuo:2014iza}
\begin{equation}
\Delta/T_H^4\sim \phi_H^2,~~~\frac{\mathd F_Q}{\mathd T_H}\sim-\phi_H,~~~~{\mathcal L}=\mathe^{-F_Q(T)/T}.
\end{equation}
The correction in the normalized trace anomaly is quadratic in $\phi_H$, while linear in the Polyakov loop. Therefore, no matter what operator is dual to the dilaton, the quadratic corrections of them in lattice data can not be implemented simultaneously. Two examples, with $\Delta_+=2,3$, are given in~\cite{Zuo:2014iza} to confirm such a conclusion. Interestingly, the results in both models show that the asymptotic expansion is actually valid in almost the whole deconfined region.
\section{Summary}
The quadratic corrections of the thermal quantities and the Polyakov loop in the deconfined phase are studied with the gauge/string duality. In the global AdS with compact boundary space, confinement is realized kinematically and quadratic corrections appear in all physical quantities. However, when confinement is generated from the dilaton, one can not achieve the quadratic terms in all of them simultaneously. This is in accordance with the fact that there is no dimension-2 gauge-invariant operator on the field theory side. These results indicate that confinement, and also quadratic corrections, must be induced by some global or non local effects of the bulk spacetime, instead of some local field.
\section*{Acknowledgments}
The work on ${\mathcal N}=4$ SYM is done in collaboration with Yi-Hong Gao. I am grateful to Pietro Colangelo, Floriana Giannuzzi and Stefano Nicotri for stimulating discussions and helpful comments. This work is partially supported by the National Natural Science Foundation of China under Grant No. 11405065, No. 11445001 and No. 11135011.

\end{document}